\documentclass[sn-mathphys-num]{sn-jnl}

\usepackage{graphicx}%
\usepackage{multirow}%
\usepackage{amsmath,amssymb,amsfonts}%
\usepackage{amsthm}%
\usepackage{mathrsfs}%
\usepackage[title]{appendix}%
\usepackage{xcolor}%
\usepackage{textcomp}%
\usepackage{manyfoot}%
\usepackage{booktabs}%
\usepackage{algorithm}%
\usepackage{algorithmicx}%
\usepackage{algpseudocode}%
\usepackage{listings}%

\theoremstyle{thmstyleone}%
\theoremstyle{thmstyletwo}%

\theoremstyle{thmstylethree}%
\begin{document}

\title[Article Title]{
Synergistic integration of physical embedding and machine learning enabling precise and reliable force field
}

\author[1,2]{\fnm{Lifeng} \sur{Xu}}
\author*[1,2]{\fnm{Jian} \sur{Jiang}}\email{jiangj@iccas.ac.cn}

\affil[1]{\orgdiv{Beijing National Laboratory for Molecular Sciences, State Key Laboratory of Polymer
Physics and Chemistry, Institute of Chemistry, Chinese Academy of Sciences}, \orgaddress{ \city{Beijing}, \postcode{100190}, \state{P.R.}, \country{China}}}

\affil[2]{\orgdiv{University of Chinese Academy of Sciences}, \orgaddress{ \city{Beijing}, \postcode{10049}, \state{P.R.}, \country{China}}}


\abstract{The machine learning force field has achieved significant strides in accurately reproducing the potential energy surface with quantum chemical accuracy. However, it still faces significant challenges, e.g., extrapolating to uncharted chemical spaces, interpreting long-range electrostatics, and mapping complex macroscopic properties. To address these issues, we advocate for a synergistic integration of physical principles and machine learning techniques within the framework of a physically informed neural network (PINN). This innovative approach involves the incorporation of physical constraints directly into the parameters of the neural network, coupled with the implementation of a global optimization strategy. We choose the AMOEBA+ force field as the physics-based model for embedding, and then train and test it using the diethylene glycol dimethyl ether (DEGDME) dataset as a case study. The results reveal a significant breakthrough in constructing a precise and noise-robust machine learning force field. Utilizing two training sets with hundreds of samples, our model exhibits remarkable generalization and DFT accuracy in describing molecular interactions and enables a precise prediction of the macroscopic properties such as diffusion coefficient with minimal cost. This work provides a crucial insight into establishing a fundamental framework of PINN.}

\maketitle

\section{Introduction}\label{sec1}

Molecular dynamics (MD) simulation is a pivotal tool for understanding the structure, property, and behavior of molecular systems, significantly advancing the frontiers of chemistry \cite{1}, biology \cite{2, 3}, and materials \cite{4} science. Central to MD simulations is the force field (FF), which encodes interatomic interactions according to the chemical environment, thereby governing the motion and state of atoms. Attributed to the rapid and robust calculation of classical FF, MD enable exploration of systems comprising millions of atoms over microsecond timescales \cite{5, 6}, providing fundamental insight into thermodynamic and dynamic properties. Nevertheless, a FF based on simple physical expression usually imposes limitations on accurately capturing complex interactions such as charge penetration and polarization effects \cite{7}, resulting in notable deviations from reality. Hence, there has been a pressing demand for high-precision force fields to ensure the faithful rendition of the structural, thermodynamics, and dynamics features of real systems.

In recent years, numerous machine learning (ML) force fields (FFs) achieving DFT accuracy in simulating interatomic potential have been proposed, e.g., DeepPot-SE \cite{8}, DimeNet \cite{9}, PaiNN \cite{10}, GemNet-T \cite{11}, NequIP \cite{12} and Allegro \cite{13} MLFFs, etc \cite{14}. The inherent training mechanism and flexible structure of neural network enable it to efficiently and accurately consider high-dimensional problems \cite{15} such as many-body interaction \cite{16}. However, in the absence of physical meaning, the traditional MLFFs are lack of extrapolative generalization, thus leading to significant model hallucination when applied to untrained data \cite{14, 17, 18, 19}. Additionally, most of MLFFs only focus on learning local atomic environment within a cutoff distance \cite{20}. The absence of long-range interaction renders MLFFs ineffective in simulating charged systems such as  electrolytes \cite{21}, and proteins \cite{22}. Fortunately, this challenge can be addressed by using a physically informed model based on the Particle Mesh Ewald (PME) technique \cite{23}. That is to say, physical knowledge can act as an inductive bias that avoids non-physical phenomena and suppresses the model hallucination.

In fact, MLFFs embedded with physical knowledge have gradually caught the attention of researchers \cite{21, 24, 25, 26}. Mishin et al. developed a physics-informed neural networks based on analytical bond-order potential, significantly improving the generalization capability of MLFFs \cite{27}. Gao and Remsing proposed a self-consistent field neural network capable of separately learning long-range and short-range interactions \cite{12}. Moreover, other strategies that integrate neural network with physical laws are also explored to construct physics-informed MLFFs. \cite{28, 29, 30}. However, accurately predicting the more intricate and far-reaching macroscopic properties from a bottom-up perspective still faces significant challenges, primarily due to the distribution shift of sampling ensembles \cite{31} and the label noise within training set \cite{18}. To tackle this obstacle, some experts endeavor to establish end-to-end mapping networks \cite{32}, substituting ensemble-averaged properties with multistate Bennett acceptance ratio (MBAR) estimators to circumvent gradient explosion \cite{33, 34}. Nevertheless, this approach is ineffective in predicting kinetic and dynamic properties such as diffusion coefficient, as well as certain physical and chemical quantities that require substantial costs to obtain by experiments \cite{35}. Therefore, although embedding physical constraint can improve MLFFs, bottom-up construction of a transferable and robust MLFF capable of accurately predicting multiscale properties remains a tremendous challenge.

In this work, we develop a PINN based on the AMOEBA+ potential, hereinafter referred to as the AMOEBA plus neural network (APNN) model. The network parameters of the APNN model are refined under physical constraints. Furthermore, we design a novel global optimizer, Tabu-Adam algorithm, to prevent the model from getting trapped in undesirable local minima. The APNN integrated with strict physical constraints and Tabu-Adam optimizer exhibits high-accuracy as well as strong generalization and robustness. To demonstrate the capabilities of our MLFF model, we train and test our model using a dataset comprised of DEGDME as an example. The results show that our model not only reproduce the microscopic physical quantities such as intra- and inter-molecular interactions with the quantum chemical accuracy, but also accurately predict the macroscopic properties measured experimentally from a bottom-up perspective. The integration of machine learning model with deep-rooted physical knowledges offers a promising technology roadmap for overcoming the inherent limitations of existing MLFF models, paving the way for more accurate and reliable predictions in molecular dynamics simulations.

\section{Results}\label{sec2}
\subsection{AMOEBA Plus Neural Network (APNN) model}\label{subsec1}
To construct a PINN with both well-established physical principles and high computational efficiency, we embed the AMOEBA+ potential \cite{36} into a neural network to build the APNN model. AMOEBA+ is regarded as one of the most cutting-edge classical force fields \cite{37, 38} to date. Its physical model not only considers many-body polarization effects but also encompasses charge penetration and charge transfer effects. Owing to the advanced iterative algorithms and parallel efficiency within the Tinker9, the AMOEBA+ force field can also achieve large-scale simulation of complex systems with affordable computational cost (e.g., simulating a system of 10,000 atoms on a GPU4090 with the speed of $\sim$30ns/day). Therefore, the corporation of AMOEBA+ and neural network is one of the most ideal choices so far.

In the APNN model, all activation functions are custom-defined, each with a clear physical meaning. The force field parameters are embedded in the activation functions across each layer and optimized during the training process. The schematic of APNN as shown in Fig. 1a. First, the cartesian coordinates and bonding information of atoms are fed into the input layer. In the second layer, the distance-dependent structural features are extracted to obtain local environment of atoms. Herein, $ \mathbf{T}_{ij} $ is the distance perception matrix between multipole moments. The expression is
\begin{equation}
\mathbf{T}_{ij} = \nabla^{k} \left( \frac{f(|\mathbf{r}_{ij}|)}{|\mathbf{r}_{ij}|} \right),\label{eq1}
\end{equation}
where $ \mathbf{r}_{ij} $ means the displacement vector from atom \textit{i} to \textit{j} and $ \mathrm{\nabla}^k $ is \textit{k}-order gradient operator, $ k = m + n $ with $m$ and $n$ denote the expansion orders of the multipole moments corresponding to atoms $i$ and $j$, respectively. For example, we set $m=0$ and $n=1$ when $ \mathbf{T}_{ij} $ refers to the distance perception matrix between the monopole of atom \textit{i} and dipole moment of atom \textit{j}. ${f(|\mathbf{r}_{ij}|)}$ is the damping function. Besides, the $ {\theta}_{ijk}$, ${\varphi}_{ijkl}$, ${\chi}_{ijkl} $ represent the angle, dihedral and abnormal dihedral. In the subsequent hidden layers, we calculate the induced dipole moments ( Eq. (10) ) and each energy term ( Eq. (5), Eq. (11-13), etc ), Finally, we sum up all interactions of the same type in the output layer. More details about $ \mathbf{T}_{ij} $ and AMOEBA+ potential are shown in the Method section.

\subsection{Tabu-Adam Algorithm}\label{subsec2}
Although the strict physical constraint endows APNN with enhanced generalization, it introduces considerable challenges to the optimization algorithm. The traditional gradient descent method, based on backpropagation neural network and Adam optimizer, has proven effective in various applications of machine learning\cite{39, 40}. However, its intrinsic local search mechanism may lead to local minima, particularly in APNN (see Fig. 3). Inspired by the niche technology, we propose an innovative metaheuristic Tabu search algorithm that perfectly integrates with the Adam optimizer, substantially augmenting its capacity for global exploration. Here, we refer to this collaborative optimization as Tabu-Adam algorithm. As shown in Fig 1b, the search engine of Tabu-Adam algorithm involves two stages: exploitation and exploration. In the exploitation, we utilize the Adam optimizer to accelerate convergence within the search space. Once the convergence is achieved, the exploration (Tabu search) is activated to escape local optima, which redirects the search towards more promising global solutions based on accumulated experience. In Tabu search, we aim to find a global point situated in a previously unexplored area, ideally far away from inferior points. To achieve it, we use the radiation intensity to represent the fitness of the point \textit{i}, which can be calculated by a surrogate model inspired by niche technique:
\begin{equation}
\mathrm{Rad}_{i}=\sum\limits_{j=1}^{N\mathrm{p}}{L({{\mathbf{p}}_{j}})\cdot D({{r}_{ij}})}+R({{\mathbf{p}}_{i}},{{\mathbf{B}}_{\min }},{{\mathbf{B}}_{\max }}).\label{eq2}
\end{equation}
On the right-hand side of Eq. (2), the first term and the second term respectively corresponds to the radiation intensities generated by the tabu list and boundary. In the first term, $ N_\mathrm{p} $ is the total number of points in previous searches, $L(\mathbf{p}_j)$ means the loss at the \textit{j}-th search, and $D(r_{ij})$ corresponds to the decay function depending on manhattan distance $r_{ij}$ between the point $i$ and $j$. In the second term, the $\mathbf{B}_{\min}$ and $\mathbf{B}_{\max}$ represent the lower and upper boundary in parameter space. The details about surrogate model can be found in the Method section. In general, radiation intensity is higher near inferior points compared to that near superior points, and also higher in already explored areas than in unexplored areas. Therefore, we can assess the fitness of arbitrarily specified point based on its negative correlation with the radiation intensity. The computational cost of evaluating radiation intensity is negligible, so we can evaluate a large number of points distributed randomly, and then filter them to obtain the optimal point. After the pre-screening, we calculate the loss function of the optimal point. If the loss is deemed acceptable, we designate this point as the initial point for the Adam search; otherwise, we continue iterating the Tabu search. In this alogrithm, Tabu search provides a macroscopic orientation for navigating the entire searching space by summarizing previous exploratory experiences. As a complement, the Adam algorithm is dedicated to meticulously probing a better solution nearby. Hence, the combination of these two distinct approaches fosters a more robust and effective solution-finding mechanism, adept at navigating complex problem spaces to unearth optimal solutions \cite{41}.

\subsection{Training framework}\label{subsec3}
Fig. 1c presents the training process of our APNN model. The entire training framework includes three aspects: preparing training data, physical refinement and model training. The left flowchart in Fig. 1c depicts the creation of training dataset including mono- and bi-molecular conformations. A framework that automatically manages the rotation-translation operator is employed for traversing the required conformations. On the right of Fig. 1c is the physical refinement framework of neural network parameters, which is achieved through different physical models. The initial parameters are obtained by poltype2\cite{42}. The equilibrium structure and force constants are refined to match optimized geometries and vibrational frequencies obtained from B3LYP-D3(BJ)/def-TZVP. The polarizabilities are corrected according to GDMA code\cite{46, 47} and atoms-in-molecules (AIM)\cite{48} analysis. Subsequently, the permanent multipoles can be derived from Eq. 15 by updating polarizabilities. The physical refinement process establishes an intuitive mapping relationship between the neural network parameters and macroscopic properties. When the above processes are successfully completed, we start training our APNN model using the physical refinement parameters as the initial model parameters. First, the range of parameters insensitive to the loss function are narrowed based on sensitivity analysis. Then, the backpropagation and Tabu-Adam algorithm are employed to train the parameters of APNN. In this process, certain constraints remain, such as penalizing for the deviation from initial parameter (Eq. S6 in Supporting information), maintaining small difference for parameters with similar chemical environments (Eq. S7 in Supporting information), and ensuring the net charge equal to the preset value (Eq. S8 in Supporting information). Furthermore, the constraint of the net charge is regarded as a hard constraint which provides positive guide for the evolution of APNN. To ensure a thorough training, the boundary tracking strategy is also employed to adaptively adjust the boundary of parameter space.

\begin{figure}[h]
\centering
\includegraphics[width=0.9\textwidth]{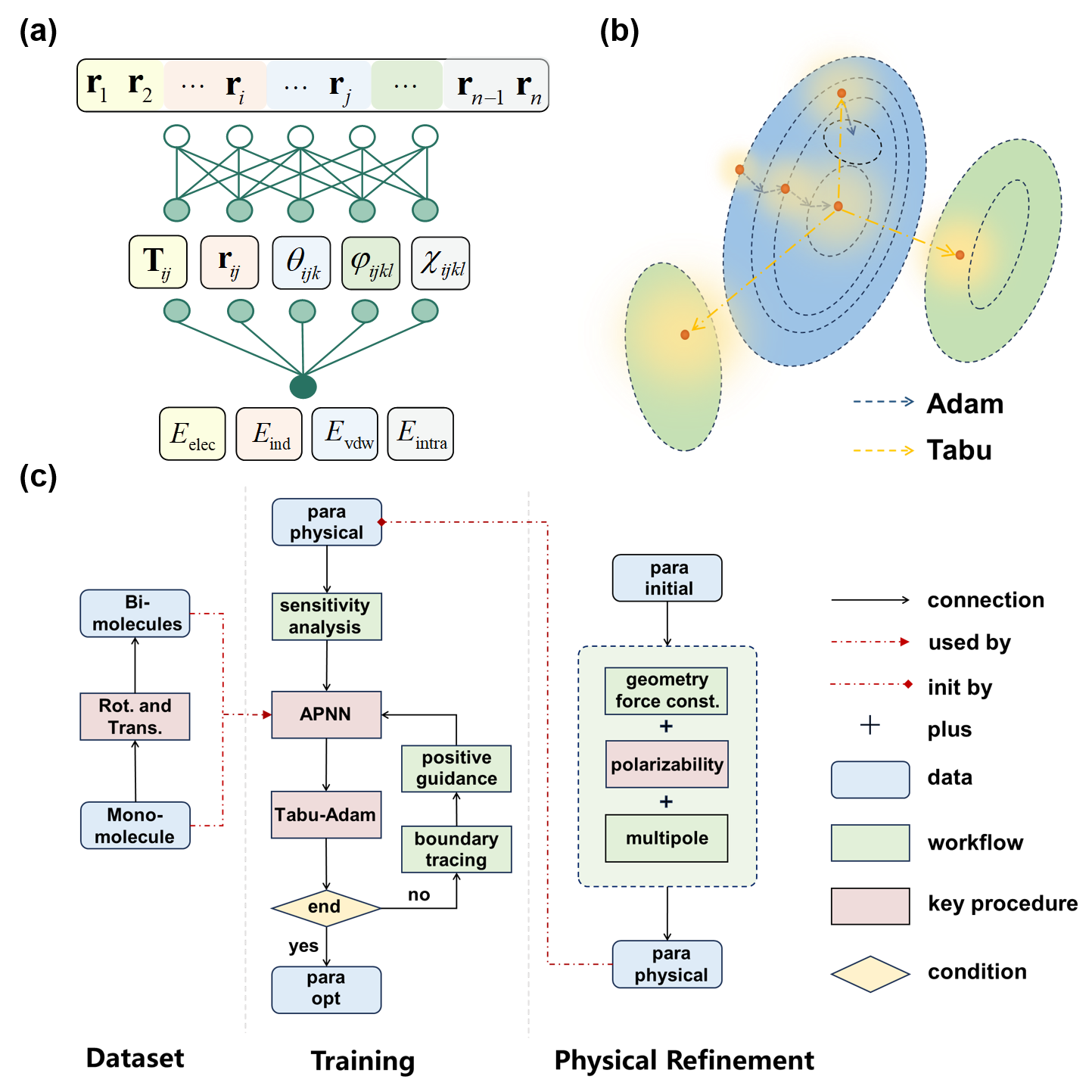}
\caption{\textbf{ \textbar Schematic of APNN, Tabu-Adam algorithm, and the training framework.} (a) Diagram of the neural network structure of APNN. (b) Schematic of Tabu-Adam algorithm, the yellow arrow indicates Tabu search (exploration), while the blue arrow corresponding to Adam search (exploitation). The yellow shading represents the radiation intensity. (c) The pipeline of whole training framework, which consists of three parts: dataset construction, physical refinement and model training.}\label{fig1}
\end{figure}

\subsection{Example: DEGDME}\label{subsec4}

Accurately simulating organic electrolyte systems remains challenge due to the imprecise description of certain interactions, such as many-body polarization, short-range charge penetration, and charge transfer. DEGDME is widely used in the fields of electrolyte due to its excellent solubility, safety, and electrochemical performance \cite{43, 44}. Here, we take DEGDME as an example to train APNN model and test its ability to describe the micro- and macro-property. In this case, we will provide a detailed explanation for all processes.

In the construction of training dataset, MD-enhanced sampling or active learning sampling are typically employed to capture molecular conformations. However, attributed to the embedding of the physical model, it’s sufficient for APNN to extrapolate the interactions across all conformational spaces relying on the primary structural features. Therefore, we do not use the typical method but develop an automatic alogrithm to construct a non-redundant dataset for APNN training. To alleviate the error compensation arising from intra- and inter-molecular interactions, we individually create mono- and bi-molecular conformations to train their corresponding interactions in our following work.
The training dataset for intramolecular interactions consists of 600 monomolecular structures, obtained by simulating annealing dynamics from 500K to 50K. The frequency distributions of bonds, angles and dihedrals within two different size of training sets are shown in Fig. S1 in Supporting information (SI). It is observed that the distribution of bonds, angles and diheadrals in small dataset, comprising 600 structures, shows no noticeable deviation from that in a larger dataset. That is to say, despite its small size, this dataset retains comprehensive topological structural information. To ensure accurate description of intramolecular energy, we employ double-hybrid density functionals, revDSD-PBEP86-D3(BJ)/aug-cc-pVTZ \cite{45}, to label the training dataset.

As shown in Fig. 2a, we construct the bimolecular conformations by automatically adjusting the relative spatial positions between two single molecules. First, we extract two molecules form the monomolecule dataset and operate on one of these two molecules through rotation and shift matrix, then adjust it constantly by series of judgments and decisions until the ideal conformation is found (see details in SI). Through this method, we obtained a training dataset composed of 442 bimolecular conformations. This method is highly designable and flexible and can evenly traverse the required conformations. To analyze the distribution of intermolecular distance, we plot the heat maps in Fig. 2 to characterize the occurrence frequencies of all atom pairs within the distance ranges of (b) 0.6$ \sim $0.7R, (c) 0.7$ \sim $0.9R, (d) 0.9$ \sim $1.1R, and (e) 1.1R $ \sim $ $\infty$, where R is the equilibrium distance of the corresponding atom pair. Figures 2(b)-(e) encompass majority of the distances between atom pairs in two molecules. Typically, the sampling of bimolecular conformations focuses on distances greater than 0.7R since extremely close intermolecular distances are regarded as rare events due to the strong repulsive interactions \cite{36}. However, we found that neglecting to train the conformations featuring the interatomic distances within the range of 0.6$ \sim $0.7R may lead to inaccurate descriptions of short-range Van der Waals energy, resulting in deviations in macroscopic properties like diffusion coefficients. As shown in Figures 2(b)-(e), in each distance range, the occurrence frequency distribution of  the atom pairs primarily concentrates within 1.5 standard deviations. It means that the occurrence frequency of each atom pair within different  distance ranges  is close to the corresponding average frequency. That is to say,  the bimolecule dataset comprising only 442 samples can rationally traverse the majority of the expected distance ranges. For a molecule with 23 atoms, the sample size of 442 is considerably smaller than the expected value (23×23×4, where 4 is the number of distance ranges). It suggests that the dataset construction strategy proposed in this work can be used to create a low-redundancy training dataset. Finally, we employ the SAPT2+/aug-cc-pVTZ method to label the intermolecular energy of bimolecular dataset and divide it into three parts: electrostatics, induction, Van der Waals (exchange + dispersion) term. This energy decomposition offers a seamless match to the energy terms in APNN model.

\begin{figure}[h]
\centering
\includegraphics[width=0.9\textwidth]{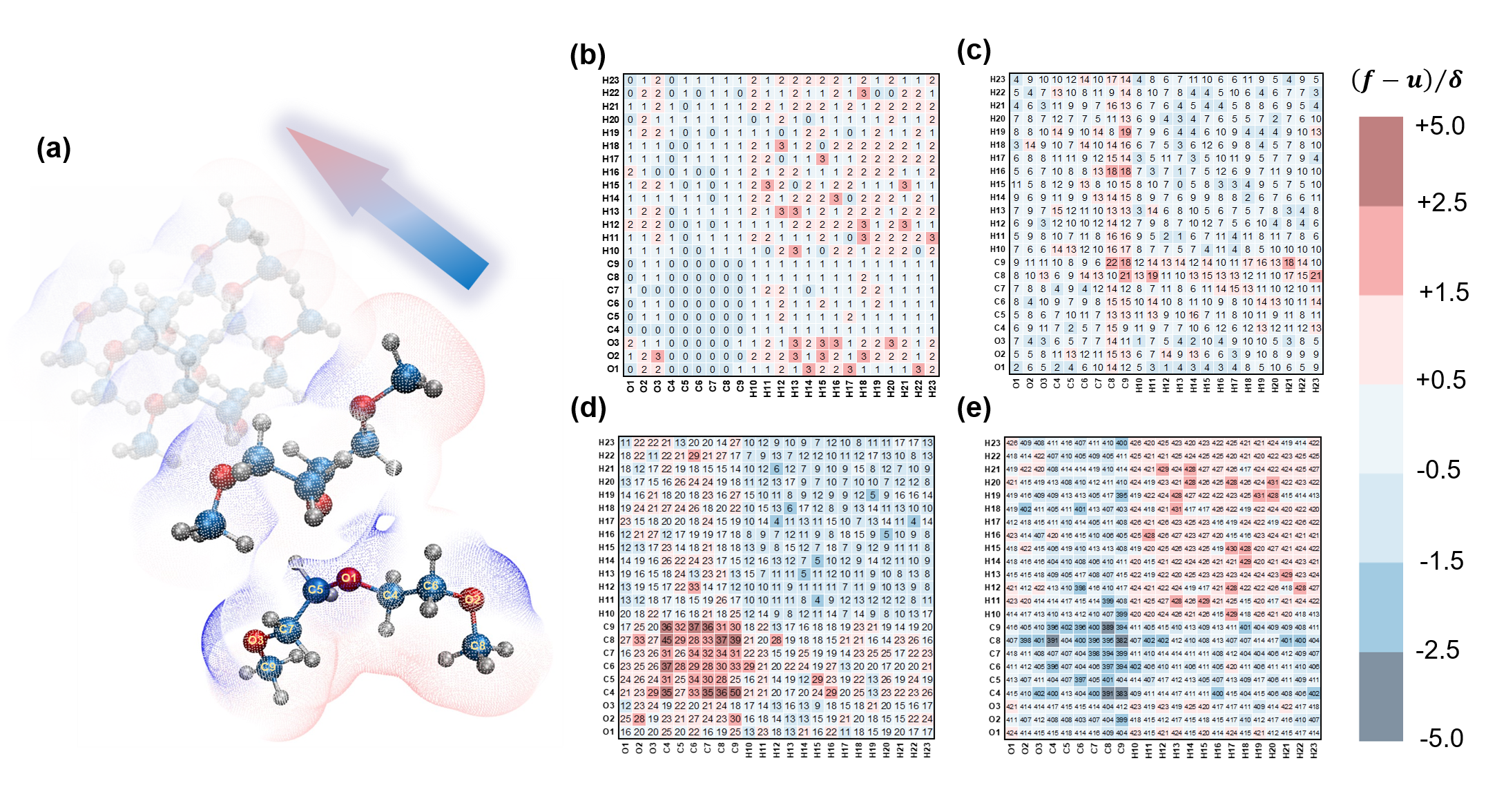}
\caption{\textbf{ \textbar Construction of training data.} (a) Schematic of automatically constructing bimolecular conformation. The occurrence frequencies of all atom pairs within the distance ranges of (b) 0.6$ \sim $0.7R, (c) 0.7$ \sim $0.9R, (d) 0.9$ \sim $1.1R, and (e) 1.1R$ \sim $+$\infty$. The row and column respectively stand for the atomic index of each molecule, and occurrence frequency of the corresponding atomic pairs is represented by the grid values. The occurrence frequency close to the mean frequency trends to be light color.}\label{fig2}
\end{figure}

In APNN model, the majority of parameters hold intrinsic physical meaning, e.g., force constant, polarizability, permanent multipole, equilibrium geometry structure and so on. These parameters establish direct mapping to the macroscopic properties. Among these parameters, polarizability serves as a direct factor determining the extent of the atomic polarization effects, crucial for precisely predicting the diffusion coefficient in electrolyte systems. The polarizability can be defined as:
\begin{equation}
a = -\lim_{\mathbf{E}\rightarrow 0} \frac{\partial \mathbf{\mu}}{\partial \mathbf{E}}.\label{eq3}
\end{equation}
where $\mathbf{\mu}$ and $\mathbf{E}$ are the dipole moment and external electric field, respectively. The polarizability can be obtained by combining GDMA \cite{46, 47} with AIM \cite{48} theory (See details of physical refinement process of polarizability and other parameters in SI). Table 1 shows the parameter values of polarizability before and after the physical refinement. The refined oxygen atom has a larger polarizability value than the carbon atom due to its lone pair electron and higher electronegativity. The terminal hydrogen atom (H\textsubscript{out}) receives less shielding from surrounding atoms, thus exhibits greater sensitivity to external electric fields. Obviously, the revised polarizability parameters align better with chemical intuition.

\begin{table}[h]
\caption{The polarizability before and after the physical refinement}\label{tab1}
\begin{tabular*}{0.68\textwidth}{@{\extracolsep\fill}p{2cm}cc}
\toprule%
&\multicolumn{2}{@{}c@{}}{Polarizability} \\
\cmidrule{2-3}%
 & Initial & Refined  \\
\midrule
O\textsubscript{in}   & 0.8370 & 0.9983\\
O\textsubscript{out}  & 0.8122 & 0.9900\\
C\textsubscript{in}   & 1.3340 & 0.9039\\
C\textsubscript{mid}  & 1.4150 & 0.8827\\
C\textsubscript{out}  & 1.6196 & 0.7075\\
H\textsubscript{in}   & 0.4960 & 0.4115\\
H\textsubscript{mid}  & 0.4960 & 0.4105\\
H\textsubscript{out}  & 0.4960 & 0.4608\\
\botrule
\end{tabular*}
\footnotetext{Note: The element subscript denotes atoms that located in different chemical environments, where the in, mid and out respectively refer to atoms in the inner layer, middle layer and outer layer.}
\end{table}

To preserve the underlying physical interpretation of the refined parameter, we train the intra- and inter-molecular interactions under stringent physical constraints, which is analogous to fine-tuning a pre-trained model for a specific downstream task. During training process, the L2 penalty function is used to penalize the deviations from refined parameters. Moreover, other physical constraints are employed to maintain the relationships between parameters. For example, the parameters with similar chemical environments should maintain minimal difference, and the net charge of each molecule should be equal to its preset value. The sensitivity analysis is also implemented to narrow down the range of parameter that are insensitive to the energy. On the other hand, the boundary tracking technology is employed to ensure a thorough optimization for parameters whose ideal value lies outside the initial boundary, as shown in the flow chart of Fig. 1c.

However, the strict physical constraints  usually leads the model to converge to an unsatisfactory local minima. To address this problem, we propose the Tabu-Adam algorithm to equip the optimizer with global search capability as mentioned above. Fig. 3 shows the variation of the lowest loss achieved thus far with respect to the number of iterations using different optimization algorithm. The efficiency of the particle swarm optimization (PSO) algorithm, relying on population information, is significantly inferior compared to that of the Adam series optimizers. The Adam optimizer is an efficient local search alogrithm benefiting to its adaptive learning rate and backpropagation mechanism. However, as shown in Fig. 3, the Tabu-Adam optimizer proposed in this work is not only high efficient but also shows strong global search capability, which can overcome the limitations of the gradient-based algorithm in APNN.

\begin{figure}[h]
\centering
\includegraphics[width=0.9\textwidth]{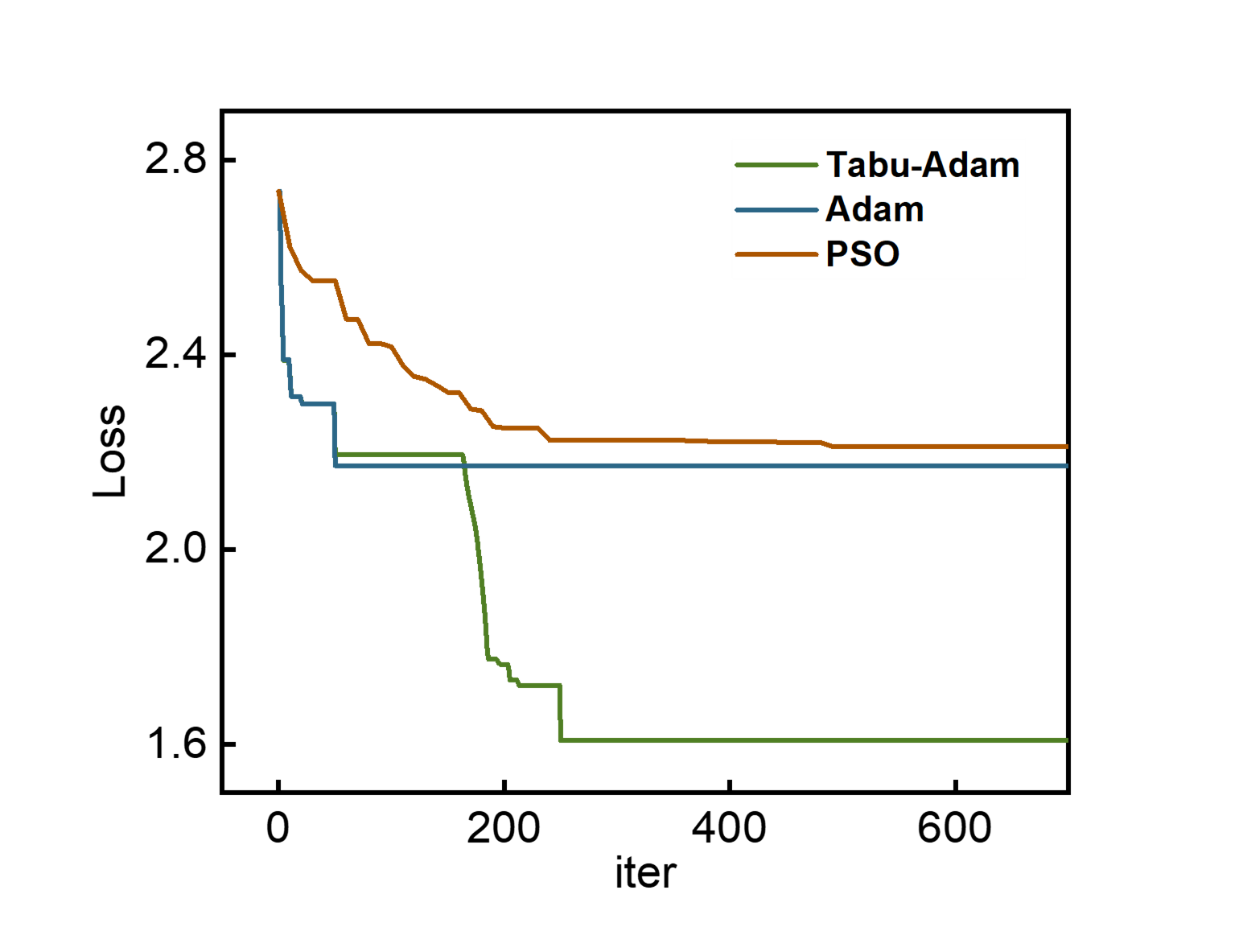}
\caption{\textbf{ \textbar The variation of lowest loss achieved thus far with respect to the number of iterations using different alogrithm.} In PSO alogrithm, the total iteration is equal to the iteration times the number of populations.}\label{fig3}
\end{figure}

The net charge constraint is a hard constraint that cannot be fulfilled solely by adding penalty term. Therefore, we continuously correct the net charge of each molecule in accordance with its preset value. Frequent correction will make the optimization inefficient, whereas sparse correction may lead the solution away from physical constraints. In this work, we correct the charge distribution every 50 steps. Fig. 4 shows the change of loss components in the lowest loss achieved thus far with respect to the number of iterations during the training of intermolecular interaction. The loss of each component changes remarkably when the verification of net charge constraint is imposed at 50 and 250 epochs. This finding suggests that the correction of net charge introduces a positive guidance steering the model towards more rational evolution.

\begin{figure}[h]
\centering
\includegraphics[width=0.9\textwidth]{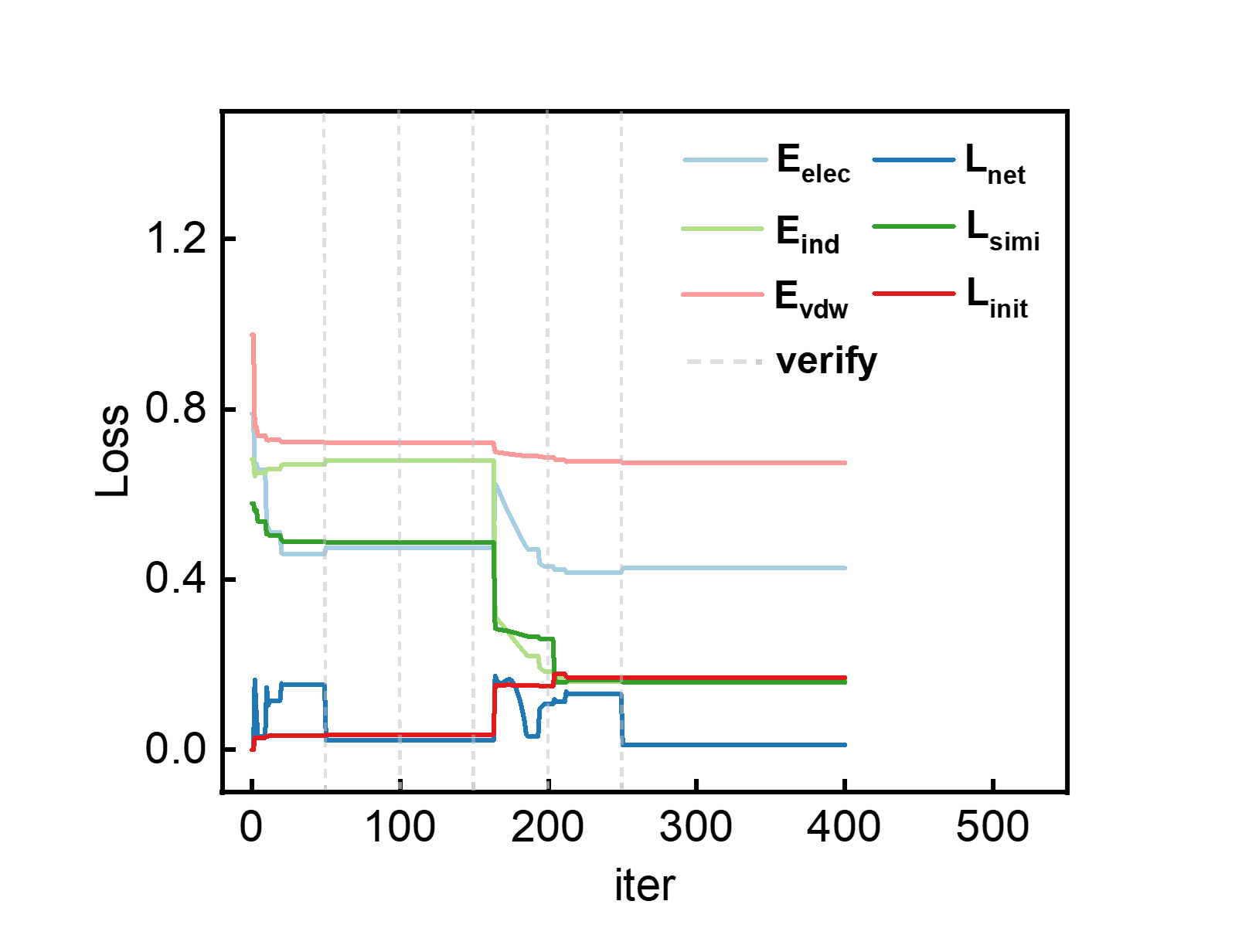}
\caption{\textbf{ \textbar The change of loss components in the lowest loss with respect to the number of iterations.} The $\mathrm{E_{elec}}$, $\mathrm{E_{ind}}$ and $\mathrm{E_{vdw}}$ signify the loss of electrostatic energy, induction energy and Van der Waals energy; $\mathrm{L_{net}}$, $\mathrm{L_{simi}}$ and $\mathrm{L_{init}}$ represent the losses of physical constraints regarding net charge, difference between parameters with similar chemical enviroments and deviation from the intial value. The Tabu-Adam algorithm is employed to train the intermolecular interactions.}\label{fig4}
\end{figure}

A strong constraint is also added to penalty deviation from the refined parameter during training intramolecular interaction. (see Fig. S2 in SI). As the equilibrium geometry structure and force constants have been corrected during physical refinement. Our focus lies primarily on the energy description of torsion and crossing stretch-bend term. After training, a slight decrease in the loss of intramolecular energy is observed, and convergence is achieved around 7000 iterations. The result is reasonable as the energy error of torsion and crossing stretch-bend term contribute only a small fraction to the loss of total intramolecular energy. Furthermore, a comparison between the energies of trained APNN and high-precision quantum mechanical method is also presented in Fig. S3 in SI, revealing a substantial agreement between them.

To validate the generalizability of the trained APNN, we further assess its energy error on a test dataset. This dataset is constructed and labeled following the same methodology applied to the training data, ensuring consistency in evaluation. The results in Fig. S4 in SI indicates that the well-trained APNN model performs exceptionally well on the test dataset as well. Furthermore, we compare the accuracy of our APNN model with commonly used ab-initio methods, sSAPT0/jun-ccpVDZ and B3LYP/6-31G(d), in describing inter- and intra-molecular energies. The quantum mechanical methods such as SAPT2+/augcc-pVTZ  and revDSD-PBEP86-D3(BJ)/aug-cc-pVTZ are employed as the benchmark. Table 2 presents the root mean square errors (RMSEs) of the results obtained by different methods on the training and test datasets. The subtle discrepancy of APNN predictions on training and test datasets again underscore the outstanding generalization of APNN force field. For predicting the intermolecular energy, the RMSE from our APNN model is remarkable lower than the one from sSAPT0/jun-ccpVDZ (0.56 vesus 1.141 kcal/mol). Moreover,  for calculating the intramolecular energy,  the RMSE from our APNN model is slightly higher than the one from B3LYP/6-31G(d) (2.94 versus 2.142 kcal/mol). These comparisons suggest that our APNN model achieves quantum chemical accuracy in predicting the microscale energies. This viewpoint is also supported by the comparison among the three methods in Fig. S5 in SI. In the meanwhile, we have to note that the APNN model possesses superior computational efficiency that surpass those of its counterpart by several orders of magnitude, particularly in large-scale systems. 

\begin{table}[h]
\caption{The RMSE of energy using APNN and medium-precision ab-initio method.}\label{tab2}
\begin{tabular*}{0.6\textwidth}{@{\extracolsep\fill}lcc}
\toprule%
&\multicolumn{2}{@{}c@{}}{RMSE to high-precision ab-initio} \\
\cmidrule{2-3}%
 & TRAIN DATA & TEST DATA  \\
\midrule
\textbf{E}\textsubscript{inter} (SAPT)    & 1.156 & 1.141\\
\textbf{E}\textsubscript{inter} (APNN)    & 0.534 & 0.560\\
\textbf{E}\textsubscript{intra} (B3LYP)   & 2.139 & 2.142\\
\textbf{E}\textsubscript{intra} (APNN)    & 2.786 & 2.960\\
\botrule
\end{tabular*}
\end{table}

Furthermore, in order to investigate the predictive accuracy of APNN for macroscopic properties, we conduct a series of simulations on bulk DEGDME at varying temperatures for calculations of the density, dielectric constant, and self-diffusion coefficient. Moreover, we compute the vibrational frequencies from equilibrium structures, which serve as a key information in infrared spectroscopy. Fig. 5 shows the comparisons between the results from the molecular simulations based on our APNN force field and the experimental measurements\cite{49, 50, 51} , revealing a great consistency between experiments and APNN predictions. As shown in Fig. 5a, we compare the results from our APNN model with experimental data over a wide temperature range. The results indicate that  the maximum relative error is less than 5\%. At certain temperatures, the APNN model nearly perfectly reproduce the experimental measurements. In predicting the dielectric constant, our APNN model shows an excellent  agreement with experiment measurement, with negligible errors observed at all temperatures except for relatively high temperatures (see Fig. 5b). In addition, the vibrational frequency obtained by our APNN model perfectly reproduce the results from B3LYP-D3(BJ)/6-31G(d) method (see Fig. 5c). This consistence is primarily due to the fitting of force constants to vibrational frequencies during the physical refinement process. Moreover, we also compare the diffusion coefficient calculated using our APNN model with the experiment data. As  illustrated in Fig. 5d, our results agree well with experiments with the maximum relatively error being less than 3×$10^{-10} m^{2}s^{-1}$. The macroscopic properties are governed by numerous microscopic behaviors, exhibiting considerable sensitivity to the microscale energy and force. Therefore, while force fields constructed using a bottom-up approach typically exhibit strong generalization and transferability, accurately predicting the macroscopic properties remains challenging. Conversely, the top-down approach exhibits the opposite characteristics. However, our APNN not only demonstrates exceptional generalization and transferability but also accurately predicts the macroscopic properties without the assistance of any experimental data and empirical parameter.  

\begin{figure}[h]
\centering
\includegraphics[width=0.9\textwidth]{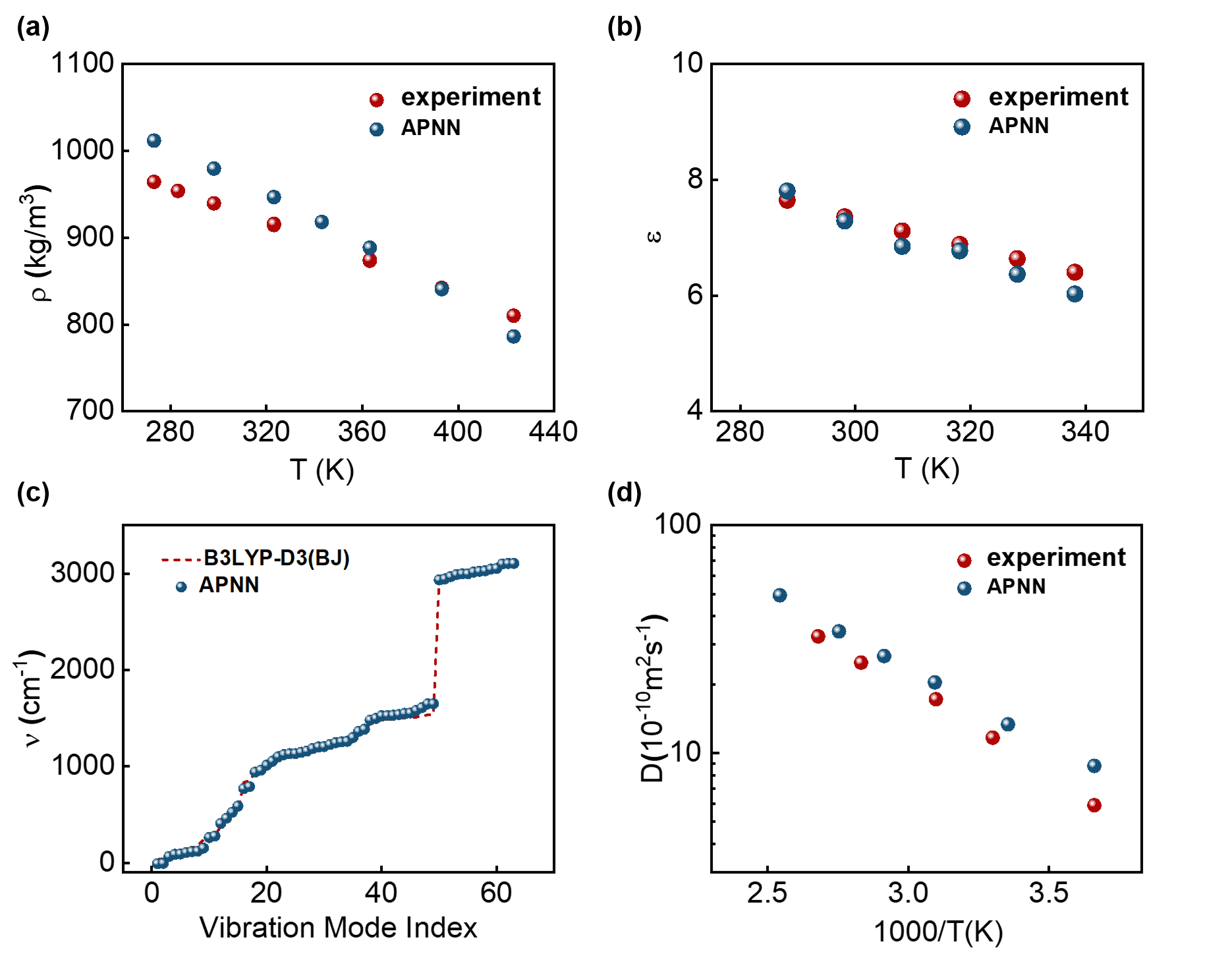}
\caption{\textbf{ \textbar Comparison of macroscopic properties between reference (experiments\cite{49, 50, 51} or DFT result) and predictions from APNN.} The (a) density, (b) dielectric constant, (c) vibration frequency and (d) self-diffusion coefficient of DEGDME are compared.}\label{fig5}
\end{figure}

\section{Conclusions}\label{sec3}

In summary, we introduce a skillful strategy to integrate physical principle and machine learning method, which endows the resulting model (APNN) with excellent generalization, robustness, and accuracy. We attribute the success of our model to the following three aspects. First, embedding the AMOEBA+ potential into neural network provides our APNN model with reliable extrapolation capability. Second, the network parameters are refined in accordance with several physical models offering more straightforward mapping to macroscopic properties. Moreover, we develop an efficient global Tabu-Adam optimizer, which greatly drives the model to the optimal solution under the stringent constraint of prior knowledge. Take DEGDME as an example, the well-trained APNN not only achieves comparable accuracy to ab-initio methods in predicting microscopic interactions such as intra- and inter-molecular energies, but also exhibits excellent agreement with experiments in predicting macroscopic properties such as density, dielectric constant, diffusion coefficient and so on. This bottom-up approach does not depend on experimental data, requiring only minimal computational resources to effectively predict both microscopic and  macroscopic properties of emerging materials. Beyond that, this method that collaboratively integrates physical constraints and machine learning is not limited to the AMOEBA+ force field, but also can extend to other physically meaningful models.

In this work, we advocate for physics-informed machine learning potentials as one of the most effective approaches to overcome the limitations of traditional force fields and non-physically embedded MLFFs. In fact, there are various ways to embed physics models into neural networks. To maximize the generalization capability of our model, we propose an “extreme” approach: impose strict physical constraints on the neural network, which both encapsulates physical principles with high fidelity and assures computational efficiency on par with traditional force fields. However, this highly customized model may result in limitation for approximating  interactions not captured by the embedded knowledge. To address this possible issue, we can replace the fuzzy part of the physical model with a flexible neural network architecture. However, balancing the flexibility of neural networks with the rigor of physical knowledge remains a significant challenge.

\section{Methods}\label{sec4}
\subsection{AMOEBA+ potential}\label{subsec5}
The total potential energy of the AMOEBA+ model consists of the bonded and nonbonded energy terms. The bonded interactions retain the formulas in AMOEBA model\cite{37}. The nonbonded interactions  (Eq. (4)) include permanent, charge transfer and polarized electrostatics as well as Van der Waals, which is the core of the polarization force field.
\begin{equation}
{{E}_{\mathrm{nonbonded}}}=E_{\mathrm{elec}}+{{E}_\mathrm{pol}}+{{E}_\mathrm{chg}}+{{E}_\mathrm{vdw}}. 	\end{equation}
The electrostatic interaction ($E_\mathrm{elec}$) is computed by multipoles truncated at quadrupoles, and is incorporated with the charge penetration effect
\begin{equation}
E_\mathrm{elec}=\sum\limits_{ij}{\frac{{{Z}_{i}}{{Z}_{j}}}{{{r}_{ij}}}+{{Z}_{i}}\mathbf{T}_{ij}^{\mathrm{damp}}{{\mathbf{M}}_{j}}+{{Z}_{j}}\mathbf{T}_{ji}^{\mathrm{damp}}{{\mathbf{M}}_{i}}+\mathbf{M}_{i}^{T}\mathbf{T}_{ij}^{\mathrm{overlap}}{{\mathbf{M}}_{j}}},               \end{equation}
where $Z_i$, $Z_j$ are nuclear charges and $\mathbf{M}_{i}$, $\mathbf{M}_{j}$ are the permanent multipole moments, the $ \mathbf{T}_{ij}^{\mathrm{damp}}$ and $ \mathbf{T}_{ij}^{\mathrm{overlap}}$ are the distance perception matrix as mentioned in Eq. (1), these two interaction tensors can be separated by different damping functions $ f^{\mathrm{damp}}(r) $ and $ f^{\mathrm{overlap}}(r)$, which are derived by Gordon et al\cite{52}:
\begin{equation}
{{f}^{\mathrm{damp}}}(r)=1-{{e}^{-{{\beta }_{i}}r}},
\end{equation}
\begin{equation}
{{f}^{\mathrm{overlap}}}\left( r \right)=1-{{e}^{-{{\beta }_{i}}r}}\frac{\beta _{j}^{2}}{\beta _{j}^{2}-\beta _{i}^{2}}-{{e}^{-{{\beta }_{i}}r}}\frac{\beta _{i}^{2}}{\beta _{i}^{2}-\beta _{j}^{2}}.
\end{equation}
In Eq. (7), the charge penetration effect is considered, where $ \beta_{i} $ and $ \beta_{j} $ are charge penetration parameters for atom \textit{i} and \textit{j}, respectively.
The induced dipole $ \mathbf{\mu}_i $ of atom \textit{i} is polarized by the total electric field: 
\begin{equation}
{{\mathbf{\mu }}_{i}}={{\alpha }_{i}}\left( \sum\limits_{j}{\mathbf{T}_{ij}^{\text{d}}{{\mathbf{M}}_{j}}}+\sum\limits_{j}{\mathbf{T}_{ij}^{\text{m}}{{\mathbf{\mu }}_{j}}} \right),
\end{equation}
where ${\alpha }_{i}$ is the polarizability of atom \textit{i}, $\mathbf{T}_{ij}^\mathrm{d} $ and $ \mathbf{T}_{ij}^\mathrm{m} $ are the interaction  for direct and mutual induction, respectively, which can be derived from the Eq. (1) with different damping functions, i.e.,
\begin{equation}
{{\mathbf{T}}_{ij}^\mathrm{m}}={{\nabla }^{2}}\left( \frac{f}{\left| {{\mathbf{r}}_{ij}} \right|} \right)=\frac{1}{4\pi {{\varepsilon }_{0}}}\left( {s}''({{r}_{ij}})\frac{3{{\mathbf{r}}_{ij}}\mathbf{r}_{ij}^{T}}{r_{ij}^{5}}-{s}'({{r}_{ij}})\frac{\mathbf{I}}{r_{ij}^{3}} \right),
\end{equation}
$s'({{r}_{ij}})$ and $s''({{r}_{ij}})$ are the first and second derivatives of the damping function with respect to ${r}_{ij}$, respectively. For mutual induction: $s'({r}_{ij})=1-e^{-au^3({r}_{ij})}$, $s''({r}_{ij})=1-(1+au^3({r}_{ij}))e^{-au^3({r}_{ij})}$; for direct induction: $s'({r}_{ij})=1-e^{-au^{3/2}({r}_{ij})}$, $s''({r}_{ij})=1-(1+1/2au^3({r}_{ij}))e^{-au^{3/2}({r}_{ij})}$, where $u(r_{ij}) = r_{ij}/(\alpha_i \alpha_j)^{1/6}$ is the scaled distance between atoms \textit{i}, \textit{j}. We can get the mapping between distance dependence matrix and dipoles by arranging the Eq. (8):
\begin{equation}
\mathbf{\mu }={{\left( \begin{matrix}
   \mathbf{\alpha }_{1}^{-1} & \mathbf{T}_{12}^{m} & \cdots  & \mathbf{T}_{1N}^{m}  \\
   \mathbf{T}_{21}^{m} & \mathbf{\alpha }_{2}^{-1} & \cdots  & \mathbf{T}_{2N}^{m}  \\
   \vdots  & \vdots  & \ddots  & \vdots   \\
   \mathbf{T}_{N1}^{m} & \mathbf{T}_{N2}^{m} & \cdots  & \mathbf{\alpha }_{NN}^{-1}  \\
\end{matrix} \right)}^{-1}}\cdot \left( \begin{matrix}
   \sum\limits_{j}{\mathbf{T}_{1j}^{d}{{\mathbf{M}}_{j}}}  \\
   \sum\limits_{j}{\mathbf{T}_{2j}^{d}{{\mathbf{M}}_{j}}}  \\
   \vdots   \\
   \sum\limits_{j}{\mathbf{T}_{Nj}^{d}{{\mathbf{M}}_{j}}}  \\
\end{matrix} \right)_.
\end{equation}
Thus, the induced dipoles can be iteratively calculated according to Eq. (10) using  the self-consistent field procedure. The many-body polarization effect in AMOEBA+ potential can be represented as:
\begin{equation}
{{E}_\mathrm{pol}}=-\frac{1}{2}\sum\limits_{ij}{{{\mathbf{\mu }}_{i}}{{\mathbf{T}}_{ij}}{{\mathbf{M}}_{j}}}.
\end{equation}
The charge transfer term is formulated as a pairwise exponential function between two atoms at near covalent distances
\begin{equation}
{{E}_\mathrm{chg}}=-\frac{1}{2}\sum\limits_{ij}{{{a}_{ij}}{{e}^{\left( -{{b}_{ij}}{{r}_{ij}} \right)}}},
\end{equation}
where $a_{ij}$ determines the magnitude of the energy associated with the transfer and $b_{ij}$ controls the decay rate. The combining rules for two atoms i and j are given as ${{a}_{ij}}=\sqrt{{{a}_{i}}{{a}_{j}}}$ and ${{b}_{ij}}=\frac{1}{2}\left( {{b}_{i}}+{{b}_{j}} \right)$.
The Van der Waals interactions capturing repulsion and dispersion effects are modeled with the buffered 14-7 equation
\begin{equation}
{{E}_\mathrm{vdw}}={{\sum\limits_{ij}{{{\varepsilon }_{ij}}\left( \frac{1.07}{{{\rho }_{ij}}+0.07} \right)}}^{7}}\left( \frac{1.12}{\rho _{ij}^{7}+0.17}-2 \right).
\end{equation}
In Eq. (13), $ \rho_{ij}={r_{ij}}/{r_{ij}^0} $ with $r_{ij}$  being the distance between atoms $i$ and $j$  and $ r_{ij}^0=({(r_{ii}^0)^3+(r_{jj}^0)^3})/({(r_{ii}^0)^2+(r_{jj}^0)^2}) $, the  $ \varepsilon_{ij} $ has the combing rule $ \varepsilon_{ij}={4\varepsilon_{ii}\varepsilon_{jj}}/({\left(\varepsilon_{ii}\right)^\frac{1}{2}+\left(\varepsilon_{jj}\right)^\frac{1}{2}}) $.

The AMOEBA+ potential has been encoded within the Tinker software, which is an efficient parallel molecular simulation tool designed for large-scale simulation. In the AMOEBA+ model, the forward propagation from atomic acoordinates to energy has already been achieved. Building upon this, we integrate a backward propagation chain to complete the loop of forward and backward propagation. By incorporating the Tabu-Adam optimizer, we proceed the model training and optimize the force field parameters informed in neural network. The advantage of this approach lies in the high degree of ease-of-use and expandability, without altering the default settings of the original AMOEBA+ force field. Moreover, the trained parameters are highly integrated with the original Tinker, facilitating rapid molecular dynamics simulations within the existing Tinker ecosystem.

\subsection{Tabu search}\label{subsec6}
In Tabu search we use the radiation intensity to define the fitness of the point \textit{i}, which can be quantitatively calculated by a surrogate model:
\begin{equation}
\mathrm{Rad}_{i}=\sum\limits_{j=1}^{N\mathrm{p}}{L({{\mathbf{p}}_{j}})\cdot D({{r}_{ij}})}+R({{\mathbf{p}}_{i}},{{\mathbf{B}}_{\min }},{{\mathbf{B}}_{\max }}),
\end{equation}
where $ N_\mathrm{p} $ is the total number of iterations in previous search, $ L(\mathbf{p}_{j}) $ and  ${\mathbf{p}}_{j}$ are the position and loss function at the \textit{j}-th search, respectively. The ${\mathbf{p}}_{i}$ is the position of the evaluated point $i$, $ r_{ij} $ is the distance between the point $i$ and $j$, and $ D(r_{ij}) $ is the decay function depend on $ r_{ij} $:
\begin{equation}
D({{r}_{ij}})=\exp (-{{{r}_{ij}}}/{{{r}_{0}}}\;),
\end{equation}
where $ r_{ij} $ is the manhattan distance presented by:
\begin{equation}
{{r}_{ij}}=\sum\limits_{k=1}^{Nd}{\left| {{p}_{i}}\left( k \right)-{{p}_{j}}\left( k \right) \right|},
\end{equation}
Where $ \mathit{p}_i\left(k\right) $ and $ {\mathit{p}_j}(k) $ denote the values of ${\mathbf{p}}_{i}$ and ${\mathbf{p}}_{j}$ in the \textit{k}-th dimension. The $N_{d}$ is the number of dimension of the parameter space. Moreover, in the Eq. (14), the $ R(\mathbf{p}_i,\mathbf{B}_{\mathrm{min}} $, $ \mathbf{B}_{\mathrm{max}}) $ represents the radiation intensity generated by the boundary:
\begin{equation}
R\left( {\mathbf{p}_{i}},{\mathbf{B}_{\min }},{\mathbf{B}_{\max }} \right)={{L}_{0}}\sum\limits_{k=1}^{N_\mathrm{d}}{\left( \left| {{B}_{\max }}(k)-{{p}_{i}}\left( k \right) \right|-\left| {{B}_{\min }}(k)-{{p}_{i}}\left( k \right) \right| \right)\cdot {{N}_\mathrm{d}}},
\end{equation}
with  ${B}_{\mathrm{min}}(k)$ and $ {B}_{\mathrm{max}}(k) $ being the vectors of the lower and upper boundary in the \textit{k}-th dimension, respectively. $L_{0}$ is the preset loss of the boundary.

\backmatter

\section*{Code availability}
The code and sample scripts will be released after review.

\section*{Data availability}
The data will be released after review.

\section*{Acknowledgements}
This work was supported by the National Key R\&D Program of China(No. 2021YFB3803200) and the National Natural Science Foundation of China under Grant No. 22273112.
\section*{Author contributions}
J.J. designed the project, J.J. and L.X. discussed the APNN architecture and Tabu-Adam alogrithm. L.X. wrote the code and performed all calculations. J.J. and L.X. wrote the manuscript.

\section*{Competing interests}
The authors declare no competing financial interests.


\bibliography{bibliography}

\end{document}